\documentstyle[12pt,axodraw]{article}
\def\beq{\begin{equation}}
\def\eeq{\end{equation}}
\def\bea{\begin{eqnarray}}
\def\eea{\end{eqnarray}}
\def\bq{\begin{quote}}    
\def\eq{\end{quote}}

\def\bq{\begin{quote}}
\def\eq{\end{quote}}

\def\a{\alpha}
\def\b{\beta}

\def\bq{\begin{quote}}
\def\eq{\end{quote}}

\def\gappeq{\mathrel{\rlap {\raise.5ex\hbox{$>$}}
{\lower.5ex\hbox{$\sim$}}}}

\def\lappeq{\mathrel{\rlap{\raise.5ex\hbox{$<$}}
{\lower.5ex\hbox{$\sim$}}}}

\def\bbz{fa Z \kern-8.9pt Z}

\begin{document}

\baselineskip 24pt
\newcommand{\sheptitle}
{Bi-Maximal Neutrino Mixing in the MSSM
with a Single Right-Handed Neutrino}

\newcommand{\shepauthor}
{S. Davidson$^1$ and S. F. King$^2$ }

\newcommand{\shepaddress}
{$^1$Theory Division, CERN, CH-1211 Geneva 23, Switzerland\\
$^2$Department of Physics and Astronomy,
University of Southampton, Southampton, SO17 1BJ, U.K.}

\newcommand{\shepabstract}
{We discuss neutrino masses in the framework of a minimal extension
of the minimal supersymmetric standard model (MSSM) consisting of 
an additional single right-handed neutrino superfield $N$
with a heavy Majorana mass $M$, which
induces a single light see-saw mass $m_{\nu_3}$
leaving two neutrinos massless at tree-level.
This trivial extension to the MSSM may account for the atomospheric
neutrino data via $\nu_{\mu}\rightarrow \nu_{\tau}$ oscillations
by assuming a near maximal mixing angle $\theta_{23}\sim \pi/4$ and 
taking $\Delta m_{23}^2 \sim m_{\nu_3}^2 \sim 2.5\times 10^{-3}\ eV^2$.
In order to account for the solar neutrino data we appeal to 
one-loop radiative corrections involving internal loops
of SUSY particles, which we show 
can naturally generate an additional light neutrino mass 
$m_{\nu_{2}}\sim 10^{-5}\ eV$ again
with near maximal mixing angle $\theta_{12}\sim \pi/4$.
The resulting scheme corresponds to so-called ``bi-maximal''
neutrino mixing involving ``just-so'' solar oscillations.}

\begin{titlepage}
\begin{flushright}
CERN-TH/98-256\\
hep-ph/9808296\\
\end{flushright}
\begin{center}
{\large{\bf \sheptitle}}
\bigskip \\ \shepauthor \\ \mbox{} \\ {\it \shepaddress} \\ \vspace{.5in}
{\bf Abstract} \bigskip \end{center} \setcounter{page}{0}
\shepabstract
\begin{flushleft}
CERN-TH/98-256\\
\today
\end{flushleft}
\end{titlepage}

Atmospheric neutrino data from Super-Kamiokande \cite{SK} 
and SOUDAN \cite{SOUDAN}, when combined with the recent CHOOZ data
\cite{CHOOZ}, are consistent with
$\nu_{\mu}\rightarrow \nu_{\tau}$ oscillations
with near maximal mixing and 
$\Delta m_{23}^2 \sim 2.5\times 10^{-3}\ eV^2$.
\footnote{The data are equally consistent if $\nu_{\tau}$ is replaced by
a light sterile neutrino, and indeed many authors have considered
adding an extra light singlet neutrino state \cite{sterile}, although
we shall not do so here.} A critical appraisal of current
neutrino data can be found in \cite{Barbieri,bahcall}.
In practice a common standard approach to neutrino masses
is to introduce three right-handed neutrinos with a heavy 
Majorana mass matrix. When the usual Dirac mass matrices of neutrinos
and charged leptons are taken into account, the see-saw mechanism then
leads to the physical light effective Majorana neutrino masses and
mixing angles relevant for experiment. 
There is a huge literature
concerning this and other kinds of approach to neutrino masses
which is impossible to do justice to.
In ref.\cite{recent} we merely list a few recent papers, 
from which which the full recent literature may be reconstructed.
For older work on neutrino masses see for instance \cite{BP},
where extensive references to earlier work may be found.

In a recent paper \cite{king} one of us
followed a minimalistic approach and introduced below the GUT scale 
only a single ``right-handed neutrino'' 
\footnote{ Since CP conjugation converts a right-handed neutrino into
a left handed anti-neutrino, it should strictly be called a gauge
singlet.}
$N$ into the standard model (or supersymmetric
standard model) with a heavy Majorana mass $\frac{1}{2}M\bar{N}N^c$ where
$M<M_{GUT}$. With only a single right-handed neutrino the low energy
neutrino spectrum consists of a light neutrino
$\nu_3 \approx s_{23}{\nu}_{\mu }+c_{23}{\nu}_{\tau }$ with mass 
$m_{\nu_3}$, plus two massless neutrinos
consisting of the orthogonal combination
$\nu_{0}\approx c_{23}{\nu}_{\mu }-s_{23}{\nu}_{\tau }$
together with $\nu_e$ which we have here assumed for simplicity
has a zero Yukawa coupling $\lambda_e$ to $N$.
\footnote{We shall use notation such as $s_{23}\equiv \sin \theta_{23}$,
$c_{23}\equiv \cos \theta_{23}$,
$t_{23}\equiv \tan \theta_{23}$, 
extensively throughout this paper.}
By contrast the Yukawa couplings $\lambda_{\mu}$ and $\lambda_{\tau}$ 
of ${\nu}_{\mu }$ and ${\nu}_{\tau }$ to $N$ are non-zero and
determine the 23 mixing angle via $t_{23}=\lambda_{\mu}/\lambda_{\tau}$.
Maximal mixing corresponds to $\lambda_{\mu}=\lambda_{\tau}$,
with $\lambda_e=0$ and $\Delta m_{23}^2=m_{\nu_3}^2$ may be chosen to
be in the correct range to account for the atmospheric neutrino data.

In order to account for the solar neutrino \cite{SNP} data via the MSW
effect \cite{MSW} we were forced to 
depart from minimality by introducing a small additional tau neutrino mass
coming from GUT scale physics
$m_{\nu_{\tau}} \sim  \frac{m_t^2}{M_{GUT}}\sim few \times 10^{-3}eV$
\cite{king}. The tau neutrino mass is clearly of the
correct order of magnitude to lift the degeneracy of the two
previously massless neutrinos by just the right amount in order
explain the solar neutrino data via the MSW effect \cite{MSW}
since $\Delta m_{12}^2\sim m_{\nu_{\tau}}^2 \sim 10^{-5} \ eV^2 $.
In addition we needed to allow $\lambda_e\neq 0$
in order to generate a small mixing angle $\theta_{12} \approx
\theta_{13}$ (where CHOOZ requires the small angle MSW solution for the higher
end of the Super-Kamiokande neutrino mass range.) In order to
account for such a tau neutrino mass perturbation we suggested that
in addition to the single right-handed neutrino $N$ there are
three ``conventional'' right-handed neutrinos with GUT scale masses which,
together with quark-lepton
unification \`{a} la minimal $SO(10)$, give rise to a hierarchy of
neutrino masses: $m_{\nu_{\tau}}:m_{\nu_{\mu}}:m_{\nu_{e}}\approx 
m_t^2:m_c^2:m_u^2$ (up to Clebsch relations, RG running effects,
and heavy Majorana textures). In this scenario the tau neutrino mass
dominates and gives the desired perturbation. The original single
right-handed neutrino $N$ is then identified as belonging to an extra
multiplet below the GUT scale.

In the present paper we wish to return to minimality by assuming
there is only a single right-handed neutrino $N$ below the GUT scale 
with {\em no} additional right-handed neutrinos
at the GUT scale. More generally we shall assume {\em no} additional
operators arising from the GUT or string scale which give rise to
neutrino masses. Our starting point is then the tree-level spectrum 
described in the second paragraph above involving one massive 
plus two massless neutrinos. Our basic observation is that
in general there is no symmetry which protects the masslessness
of the two neutrinos, so they  are expected to acquire
masses beyond the tree-level. To be definite we shall show that 
masses for $\nu_{0}$ and $\nu_e$ can be generated from 
one-loop SUSY corrections and lead to a
neutrino mass $m_{\nu_2}\sim 10^{-5}$ eV, corresponding to
$\nu_2 \approx  s_{12}{\nu}_{e }+c_{12}{\nu}_{0 }$, while
the orthogonal combination
$\nu_1 \approx  c_{12}{\nu}_{e }-s_{12}{\nu}_{0 }$  
remains approximately massless. The value of the 12 mixing angle
is controlled by a ratio of soft flavour-violating parameters, 
$t_{12}=\Delta_{13}/\Delta_{23}$ which could be large,
and $\Delta m_{12}^2\sim m_{\nu_2}^2$.
In this case we arrive at approximate
``bi-maximal'' flavour mixing in both the $(\nu_e - \nu_0)$
and $(\nu_{\mu} - \nu_{\tau})$ sectors \cite{bimaximal1}. 
This may be compared to bi-maximal mixing in the 
$(\nu_e - \nu_{\mu})$ and $(\nu_{\mu} - \nu_{\tau})$ sectors
\cite{bimaximal2}. 
Both forms of bi-maximal mixing rely on
the ``just-so'' \cite{justso1,justso2} vacuum oscillation
description of the solar neutrino data. 
For example a recent best-fit
to solar neutrino data in the vacuum oscillation region 
requires \cite{bahcall}
$\Delta m_{12}^2 \simeq 6.5\times 10^{-11} \ eV^2$ 
and a mixing angle $\sin^2 2 \theta_{12}\sim 0.75$ .
Following \cite{GGR2}, we assume that
this is not in conflict with the spectrum
of electron neutrinos detected from SN1987A \cite{GGR}.
The main point of this paper is to show
that bi-maximal mixing can arise in a natural way from a bare
minimum of ingredients: the addition of a single
right-handed neutrino to the MSSM.

The MSSM plus one heavy right-handed neutrino $N$ has a superpotential
which contains, in the addition to the usual terms 
such as $\mu H_1H_2$ plus the quark and charged lepton Yukawa
couplings, the following new terms involving the new $N$ superfield,
\beq
W_{new} = \lambda_{\alpha} L_{\alpha} N H_2 + \frac{1}{2} M NN
\label{W}
\eeq
where the subscript $\alpha=e,\mu ,\tau$ indicates that we are 
in the charged lepton mass eigenstate basis, e.g. $L_e=(\nu_{eL} , e^-_L)$
where $e^-$ is the electron mass eigenstate and $\nu_e$ is the associated
neutrino weak eigenstate, and $\lambda_e$ is a Yukawa coupling to the
singlet $N$ in this basis. In this basis the soft terms involving
sneutrinos are
\bea
V_{soft} & = &
m_{{\tilde L}_\a{\tilde L}_\b^*}^2 \tilde{L}_\a \tilde{L}_\b^* 
+ m_{\tilde N \tilde{N}^*}^2 \tilde{N}  \tilde{N}^* \nonumber \\
& + &  ( \lambda_\a A_\a H_2  \tilde{L}_\a \tilde{N} +
M B_N  \tilde{N} \tilde{N} + h.c.)
\eea
The potential involving sneutrinos, including F-terms, D-terms and
soft terms is, in the usual notation where
$\tan \beta =v_2/v_1$ is the ratio of Higgs vacuum expectation values,
\bea
V & = & (m_{{\tilde L}_\a{\tilde L}_\b^*}^2 +\lambda_\a \lambda_\b v_2^2 +
\frac{1}{2}m_Z^2\cos 2\beta \delta_{\a \b})
\tilde{\nu}_\a \tilde{\nu}_\b^* +  ({M}^2 + m_{\tilde N \tilde{N}^*}^2) 
\tilde{N}  \tilde{N}^* \nonumber \\
& + & 
[(v_2A_\a \lambda_\a - \lambda_\a \mu v_2 \cot \beta)\tilde{\nu}_\a \tilde{N} 
+(\lambda_\a v_2 M)\tilde{\nu}_i \tilde{N}^* 
+ M B_N  \tilde{N} \tilde{N}+ h.c.]
\label{Lmass}
\eea
We neglect $ m_{\tilde N \tilde{N}^*}^2$ with respect to
$M^2$  for the remainder of the paper. 

The tree-level neutrino spectrum was discussed in ref.\cite{king}
and is summarised below. The usual fermionic see-saw mechanism 
which is responsible for the light effective Majorana mass 
$m_{{\nu}_\a{\nu}_\b} {\nu}_\a {\nu}_\b $
can be represented by the diagram in Fig.1.

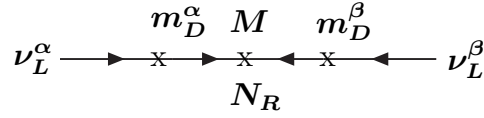
\begin{figure}[ht]
\unitlength.5mm
\SetScale{1.418}
\begin{boldmath}
\begin{center}
\begin{picture}(120,40)(0,0)
\ArrowLine(0,20)(30,20)
\ArrowLine(30,20)(50,20)
\ArrowLine(70,20)(50,20)
\ArrowLine(100,20)(70,20)
\Text(-2,20)[r]{$\nu_L^{\alpha}$}
\Text(45,10)[l]{$N_R$}
\Text(103,20)[l]{$\nu_L^{\beta}$}
\Text(24,20)[l]{x}
\Text(47,20)[l]{x}
\Text(69,20)[l]{x}
\Text(24,30)[l]{$m_D^{\alpha}$}
\Text(45,30)[l]{$M$}
\Text(68,30)[l]{$m_D^{\beta}$}
\end{picture}
\end{center}
\end{boldmath}
\caption{\small A diagrammatic representation of the see-saw mechanism. 
$M$ is the singlet Majorana mass, and we have defined 
$m_D^{\alpha} \equiv \lambda_\a v_2$. }
\label{fm23e}
\end{figure}

%\begin{itemize}
%\item rightgoing  ${\nu_\a}$, $\lambda_\a v_2$,
%rightgoing  ${N}$, Majorana mass
%insertion $  M$ , leftgoing  ${N}$,
%$\lambda_\b v_2$, leftgoing  ${\nu_\b}$. 
%DO WE KEEP THIS?
%\end{itemize}

The see-saw mechanism represented by Fig.1 leads to
the light effective Majorana matrix in the
$\nu_{e L} , \nu_{\mu L}, \nu_{\tau L}$ basis:
\beq
m_{{\nu}_\a{\nu}_\b} =
\left( \begin{array}{lll}
 \lambda_e^2 &  \lambda_e \lambda_{\mu} & \lambda_e \lambda_{\tau}     \\
\lambda_e \lambda_{\mu} & \lambda_{\mu}^2 & \lambda_{\mu} \lambda_{\tau} \\
\lambda_e \lambda_{\tau} & \lambda_{\mu} \lambda_{\tau} & \lambda_{\tau}^2
\end{array}
\right)\frac{v_2^2}{M}
\label{Maj}
\eeq
with phase choices such that the Yukawa couplings are real.
The matrix in Eq.\ref{Maj} has two zero eigenvalues and one non-zero
eigenvalue $m_{\nu_3}=\sum_{\a}|\lambda_{\a}|^2v_2^2/M$ corresponding to the
eigenvector $\nu_3=\lambda_{\a}\nu_{\a}/\sqrt{\sum_{\b}|\lambda_{\b}|^2}$.
This can easily be understood from Eq.\ref{W} where it is clear that
only the combination
$\nu_3$ couples to $N$ with a Yukawa coupling 
$\lambda_3=\sqrt{\lambda_{\b}^2}$.
In the $\lambda_e=0$ limit $\nu_e$ is massless and the other two
eigenvectors are simply
\beq
\left(
\begin{array}{l}
\nu_0 \\
\nu_3
\end{array}
\right)
=
\left(
\begin{array}{ll}
c_{23} & -s_{23}\\
s_{23} & c_{23}
\end{array}
\right)
\left(
\begin{array}{l}
\nu_{\mu} \\
\nu_{\tau}
\end{array}
\right)
\label{bbasis}
\eeq
where $t_{23}=\lambda_{\mu}/\lambda_{\tau}$ and
the $\nu_0$ is massless at tree-level.
If we allow a small but non-zero $\lambda_e$ then we denote the
massless states by primes. In this case
the weak eigenstates are related to the mass eigenstates
by \cite{king}
\beq
\left(
\begin{array}{l}
\nu_e  \\
\nu_\mu  \\
\nu_\tau 
\end{array}
\right)_L = U_0 
\left(
\begin{array}{l}
\nu'_e \\
\nu'_0 \\
\nu_3
\end{array}
\right)_L
\label{defns}
\eeq
where
\beq
U_0 = \left( \begin{array}{ccc}
 1 &  \left( \frac{c_{23}}{s_{23}}\right) \theta_1   & \theta_1     \\
-\frac{\theta_1}{s_{23}}   & c_{23} &  s_{23}   \\
0   &  -s_{23}   & c_{23} 
\end{array}
\right)
\label{U0}
\eeq
where $\theta_1 \approx
\lambda_e/\sqrt{\lambda_{\mu}^2+\lambda_{\tau}^2}$.
The unitary matrix $U_0$ is the tree-level
neutrino mixing matrix (the analogue of
the CKM matrix). This follows since the
charged weak currents are given by:
\beq
W_\mu^- ( \bar{e}, \bar{\mu}, \bar{\tau} )_L \gamma^\mu 
\left( \begin{array}{l} 
\nu_e\\
\nu_\mu\\
\nu_\tau
\end{array}
\right)_L
+ h.c.
\eeq
where $\nu_{eL}, \nu_{\mu L}, \nu_{\tau L}$ 
are neutrino weak eigenstates which couple with unit 
strength to $e$, $\mu$, $\tau$, respectively.
Note that due to the massless degeneracy 
of $\nu_0$ and $\nu_e$ the choice of basis is not unique.

The MSSM generalisation of the see-saw mechanism 
was recently discussed by Grossman and Haber \cite{GH}.
The SUSY analogue of the Majorana neutrino mass
$m_{{\nu}_\a{\nu}_\b} \bar{{\nu}_\a}^c {\nu}_\b $ is the
``Majorana'' sneutrino mass
$m_{{\tilde \nu}_\a {\tilde \nu}_\b }^2 \tilde{\nu}_\a
\tilde{\nu}_\b $. Such masses, which perhaps should more properly
be referred to as lepton number violating masses,
do not appear directly in the potential in Eq.\ref{Lmass},
but can be generated by the scalar analogue of the
see-saw mechanism. One way \cite{GH} to understand the origin of 
a ``Majorana'' sneutrino mass is to separate each sneutrino
into real and imaginary components 
$\tilde{\nu} = 2^{-1/2}(\tilde{\nu}_R +i  \tilde{\nu}_I)$ 
and $\tilde{N} = 2^{-1/2}(\tilde{N}_R + i \tilde{N}_I)$
then observe that a lepton number conserving (``Dirac'')
sneutrino mass is the same for real and imaginary
parts of the field:
\beq
m^2_{ \tilde{\nu} \tilde{\nu}^*} \tilde{\nu} \tilde{\nu}^* = 
\frac{m^2_{ \tilde{\nu} \tilde{\nu}^*}}{2} 
(\tilde{\nu}_R^2 +  \tilde{\nu}_I^2)
\eeq
whereas a lepton number violating (``Majorana'') mass contributes with
opposite sign to the  mass of the real and
imaginary components:
\beq
(m^2_{\tilde{\nu} \tilde{\nu}} \tilde{\nu} \tilde{\nu}+ h.c.)
 = m^2_{\tilde{\nu} \tilde{\nu}} (\tilde{\nu}_R^2 -  \tilde{\nu}_I^2)
\eeq
Grossman and Haber studied the $4 \times 4$
matrix for the one-family case and showed that it 
separates into two independent $2 \times 2$ matrices corresponding to
the CP even and CP odd states. To calculate the sneutrino Majorana mass
they find eigenvalues of the $2 \times 2$ matrices and take mass
differences. However one can equally well think of the Majorana
sneutrino masses diagramatically,
as the scalar analogue of the diagrammatic representation of the usual
fermionic see-saw mechanism discussed above, as follows.

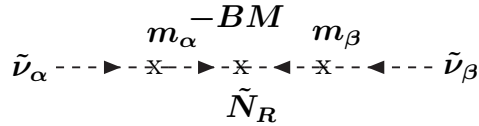
\begin{figure}[ht]
\unitlength.5mm
\SetScale{1.418}
\begin{boldmath}
\begin{center}
\begin{picture}(120,40)(0,0)
\DashArrowLine(0,20)(30,20){2}
\DashArrowLine(30,20)(50,20){2}
\DashArrowLine(70,20)(50,20){2}
\DashArrowLine(100,20)(70,20){2}
\Text(-2,20)[r]{$\tilde{\nu}_{\alpha}$}
\Text(45,10)[l]{$\tilde{N}_R$}
\Text(103,20)[l]{$\tilde{\nu}_\beta$}
\Text(24,20)[l]{x}
\Text(47,20)[l]{x}
\Text(69,20)[l]{x}
\Text(24,28)[l]{$m_{\alpha}$}
\Text(35,33)[l]{$-BM$}
\Text(68,28)[l]{$m_{\beta}$}
\end{picture}
\end{center}
\end{boldmath}
\caption{\small A scalar see-saw diagram. We have defined
$m_{\alpha} \equiv \lambda_{\alpha} v_2 M $.}
\label{f2}
\end{figure}

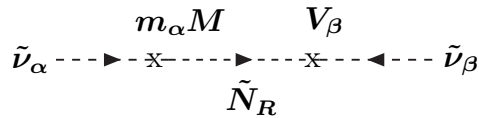
\begin{figure}[ht]
\unitlength.5mm
\SetScale{1.418}
\begin{boldmath}
\begin{center}
\begin{picture}(120,40)(0,0)
\DashArrowLine(0,20)(30,20){2}
\DashArrowLine(30,20)(70,20){2}
\DashArrowLine(100,20)(70,20){2}
\Text(-2,20)[r]{$\tilde{\nu}_{\alpha}$}
\Text(45,10)[l]{$\tilde{N}_R$}
\Text(103,20)[l]{$\tilde{\nu}_\beta$}
\Text(24,20)[l]{x}
\Text(66,20)[l]{x}
\Text(21,30)[l]{$m_{\alpha}M$}
\Text(66,30)[l]{$V_{\beta}$}
\end{picture}
\end{center}
\end{boldmath}
\caption{\small Another scalar see-saw diagram. We have defined
$V_{\beta} \equiv v_2 A_{\beta} \lambda_{\beta} - \lambda_{\beta}
\mu v_2 \cot \beta$.}
\label{f3}
\end{figure}

As in the fermionic see-saw mechanism the light effective
Majorana sneutrino masses come from integrating out a heavy state, in this
case $\tilde{N}$ which
has the lepton number violating interactions. 
Thus the scalar  see-saw diagrams involve an $\tilde{N}$
propagator. To get a contribution to 
$m_{{\tilde \nu}_\a{\tilde \nu}_\b }^2 \tilde{\nu}_\a \tilde{\nu}_\b $
that is $O(1/M)$, we need at least one power
of $M$ on top to cancel the $1/M^2$ from
the propagator. There are two possible diagrams,
illustrated in Figs. \ref{f2} and \ref{f3}.

We estimate the contribution from Fig.\ref{f2}
to be $ \simeq -B{ \lambda_\a \lambda_\b v_2^2}/{M}$,
and from Fig. \ref{f3} to be
$\simeq (A_\b  - \mu \cot \beta)\lambda_\a \lambda_\b  {v_2^2}/{M}$.
The total contribution to the light effective 
Majorana (= lepton number violating) sneutrino mass is then
\beq
m_{{\tilde \nu}_\a{\tilde \nu}_\b }^2 \simeq
(A_\b - \mu \cot \beta-B) m_{{\nu}_\a{\nu}_\b}
\label{sneutrino}
\eeq
which consists of a product
of SUSY masses times the tree-level Majorana neutrino mass.
This result agrees with the result in ref.\cite{GH}.

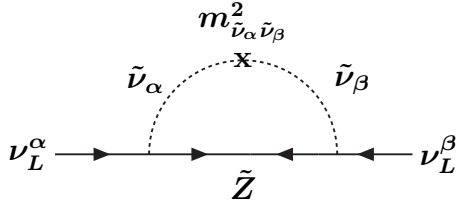
\begin{figure}[ht]
\unitlength.5mm
\SetScale{1.418}
\begin{boldmath}
\begin{center}
\begin{picture}(120,40)(0,0)
\ArrowLine(0,0)(25,0)
\ArrowLine(25,0)(52,0)
\ArrowLine(70,0)(52,0)
\ArrowLine(95,0)(70,0)
\DashCArc(50,0)(25,0,180){1}
\Text(-2,0)[r]{$\nu_L^{\alpha}$}
\Text(97,0)[l]{$\nu_L^{\beta}$}
\Text(50,25)[c]{{\bf x}}
\Text(24,20)[c]{$\tilde{\nu}_{\alpha}$}
\Text(79,20)[c]{$\tilde{\nu}_{\beta}$}
\Text(50,35)[c]{$m^2_{\tilde{\nu}_{\a} \tilde{\nu}_{\b}}$}
\Text(50,-8)[c]{$\tilde{Z}$}
\end{picture}
\end{center}
\end{boldmath}
\vspace{10mm}
\caption{\small One-loop diagram generating  a Majorana neutrino
mass. The lepton number violation comes from the
effective sneutrino ``Majorana'' (lepton number violating)
mass $m^2_{\tilde{\nu}_\a \tilde{\nu}_\b}$ arising from Figs.\ref{f2},\ref{f3}
which we have condensed into a single ``x'' here. }
\label{f4}
\end{figure}

Having generated a {\em sneutrino} Majorana mass it is straightforward to
see that such a mass will lead to one-loop radiative corrections to 
{\em neutrino} Majorana masses \cite{GH} 
via the self-energy diagram in Fig. \ref{f4}. 

The diagram involves
an internal loop
of neutralinos and sneutrinos, with the lepton number
violating Majorana sneutrino mass at the heart of the diagram.
This diagram applies to an arbitrary basis.
In particular it applies to the basis in which the tree-level
Majorana neutrino masses are diagonal, which is the most convenient basis
for calculating loop corrections to neutrino masses.
In this basis the Majorana sneutrino masses are also diagonal
(assuming that $A_{\a} \lambda_{\a}$ is aligned with $\lambda_{\a}$),
as is clear from Eq.\ref{sneutrino}, and consist of two zero Majorana
mass sneutrinos 
\footnote{The usual lepton number conserving sneutrino masses are
of course non-zero.}
plus a non-zero
Majorana mass given by Eq.\ref{sneutrino} in this basis:
\beq
m_{{\tilde \nu}_3{\tilde \nu}_3}^2 \approx 
(A_3 - \mu \cot \beta - B) m_{{\nu}_3}
\eeq
where $A_3$ is the trilinear soft parameter associated with the
Yukawa coupling eigenvalue $\lambda_3$ (ignoring flavour violation.)
According to Grossman and Haber \cite{GH} the one-loop 
correction to the tree-level neutrino mass is given by:
\beq
\delta m_{{\nu}_3}\approx \frac{g^2}{32\pi^2 \cos ^2\theta_W}
\frac{m_{{\tilde \nu}_3{\tilde \nu}_3}^2 }{\bar{m}}
\sum_j f(y_j)|Z_{jZ}|^2\equiv \epsilon m_{{\nu}_3} 
\eeq
where the $y_j={\bar{m}}^2/m_{\tilde{\chi}^0_j}^2$,  
${\tilde{\chi}^0_j}$ is the j-th neutralino, the function
$f(y_j)\sim 0.25-0.57$ and $\bar{m}^2$ is an
average sneutrino mass (see Eq.\ref{1a}, \ref{1b}, and
\ref{massins}.)
We have defined a quantity $\epsilon$ whose value is typically
$\epsilon \sim 10^{-3}$, assuming that all soft SUSY breaking
parameters and $\mu$
(including those associated with the right-handed
neutrino $N$) are of a similar order of magnitude.
Grossman and Haber actually consider the case that $B\gg 1$ TeV
but we shall assume here that $B \sim $1 TeV. Thus we conclude that
the correction to the tree-level neutrino mass is of order
0.1\% and so is utterly negligible.

Since $\delta m_{{\nu}_3}\propto m_{{\nu}_3}$ 
one might be tempted to conclude that the two neutrino states which are 
massless at tree-level remain massless at one-loop.
However this conclusion would be incorrect due to the effects
of flavour violation which we have so far ignored.
There are two possible origins of
flavour violation that can be relevant for neutrino masses:

(i) Soft lepton number conserving sneutrino masses which are
flavour-violating;

(ii) Trilinear parameters $A_\b$ which are misaligned with
the Yukawa couplings $\lambda_\b$.

In the present paper we shall focus on case (i) only
since we shall see later that the mechanism responsible for
generating small non-zero neutrino masses has two possible sources:
(i) or (i)+(ii). There is no significant 
source of small neutrino masses coming
from (ii) alone, and so for simplicity we shall ignore the effect of (ii)
in making our estimates.

We shall need to be able to move from one basis to another
and be able to deal with flavour-violating effects in any given basis.
The tool for doing this which we shall use is the
``mass-insertion approximation''.
According to the mass insertion approach
in changing basis we must rotate the superfield as a whole,
so that the gauge couplings do not violate flavour. 
This means that in the fermion mass eigenstate basis the sfermions
have off-diagonal masses, and flavour-violation is dealt with
by sfermion propagator mass insertions. To begin with we shall develop the
mass-insertion approximation in the charged lepton mass eigenstate
basis, then change basis to the tree-level neutrino mass eigenstate
basis where we shall actually perform our estimates.

In the charged lepton mass eigenstate basis for simplicity we
assume that the lepton number conserving sneutrino masses
are approximately proportional to the unit matrix:
\beq
M_{{\tilde L}_\a{\tilde L}_\b^*}^2 =
m_{{\tilde L}_\a{\tilde L}_\b^*}^2 
+ \lambda_{\alpha} \lambda_{\beta} v_2^2 + \frac{1}{2} m_Z^2 \cos 2 \beta
\delta_{\alpha \beta}
= \bar{m}^2 (\delta_{\a \b} + \Delta_{\a \b})
\label{1a}
\eeq 
where the $\Delta_{\a \b}$ are small. Note that
$\Delta$ is dimensionless. 
The inverse propagator in this basis is given by
\beq
p^2 -  \bar{m}^2 (\delta_{\a \b} + \Delta_{\a \b}) = 
(p^2 -  \bar{m}^2) \left( \delta_{\a \b} - \frac{\bar{m}^2 \Delta_{\a \b}}
{p^2 -  \bar{m}^2} \right)
\label{1b}
\eeq
so to linear order in $\Delta$, the propagator
becomes
\beq
 \frac{\delta_{\a \b}}
{p^2 - \bar{m}^2}
+ \frac{1}{p^2 - \bar{m}^2} \bar{m}^2 \Delta_{\a \b} \frac{1}{p^2 - \bar{m}^2}
\label{massins}
\eeq
where $\bar{m}$ is some average sneutrino (lepton number conserving) mass.

Now we need to relate the lepton number conserving
sneutrino mass matrix in the charged lepton
mass eigenstate basis to that in the tree-level neutrino mass
eigenstate basis. Since in the mass insertion approach we rotate each
component of the superfields equally, the rotation on the sneutrinos
is the same as that for the neutrinos and is given by the unitary
matrix $U_0$ in Eq.\ref{U0}. Thus the relation between the sneutrinos 
in the two bases is the scalar  version of Eq.\ref{defns}
\beq
\left(
\begin{array}{l}
\tilde{\nu}_e  \\
\tilde{\nu}_\mu  \\
\tilde{\nu}_\tau 
\end{array}
\right) = U_0 
\left(
\begin{array}{l}
\tilde{\nu}'_e \\
\tilde{\nu}'_0 \\
\tilde{\nu}_3
\end{array}
\right)
\label{sdefns}
\eeq
(Of course the rotation acts on complete $SU(2)_L$ doublets 
even though we only exhibit it for the neutrinos and sneutrinos.)
The relation between the soft masses in the two bases is then given by:
\beq
(\tilde{\nu}_e^*, \tilde{\nu}_\mu^*, \tilde{\nu}_\tau^*)
M_{{\tilde L}_\a{\tilde L}_\b^*}^2 
\left( \begin{array}{l} 
\tilde{\nu}_e  \\
\tilde{\nu}_\mu  \\
\tilde{\nu}_\tau 
\end{array}
\right)
=
(\tilde{\nu '}^*_e, \tilde{\nu '}^*_0, \tilde{\nu}^*_3)
{M'}_{{\tilde L}_i{\tilde L}_j^*}^2 
\left(
\begin{array}{l}
\tilde{\nu}'_e \\
\tilde{\nu}'_0 \\
\tilde{\nu}_3
\end{array}
\right)
\eeq
where the soft masses are related by
\beq
{M'}_{{\tilde L}_i{\tilde L}_j^*}^2 =U_0^{\dag}
M_{{\tilde L}_\a{\tilde L}_\b^*}^2 U_0
\eeq
Thus the soft masses in the tree-level neutrino mass eigenstate basis 
can be written as
\beq
{M'}_{{\tilde L}_i{\tilde L}_j^*}^2 
= \bar{m}^2 (\delta_{ij} - \Delta'_{ij})
\eeq 
where the $\Delta$'s are related by
\beq
\Delta'_{ij}=U_0^{\dag}\Delta_{\a \b}U_0
\label{Deltarelation}
\eeq

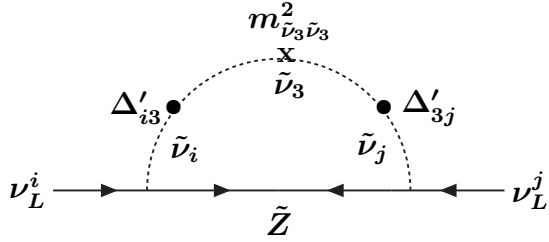
\begin{figure}[ht]
\unitlength.5mm
\SetScale{1.418}
\begin{boldmath}
\begin{center}
\begin{picture}(120,40)(0,0)
\ArrowLine(0,0)(30,0)
\ArrowLine(30,0)(60,0)
\ArrowLine(120,0)(90,0)
\ArrowLine(90,0)(60,0)
\DashCArc(60,0)(35,0,180){1}
\Text(-2,0)[r]{$\nu_L^{i}$}
\Text(122,0)[l]{$\nu_L^{j}$}
\Text(62,36)[c]{{\bf x}}
\Text(32,22)[c]{$\bullet$}
\Text(22,22)[c]{$\Delta_{i3}'$}
\Text(88,22)[c]{$\bullet$}
\Text(100,22)[c]{$\Delta_{3j}'$}
\Text(35,11)[c]{$\tilde{\nu}_{i}$}
\Text(85,11)[c]{$\tilde{\nu}_{j}$}
\Text(63,45)[c]{$m^2_{\tilde{\nu}_3 \tilde{\nu}_3}$}
\Text(63,28)[c]{$\tilde{\nu}_3$}
\Text(60,-8)[c]{$\tilde{Z}$}
\end{picture}
\end{center}
\end{boldmath}
\vspace{10mm}
\caption{ \small One-loop diagram generating  a Majorana neutrino
mass that is not aligned with
the tree level mass.  The lepton number violation comes from the
effective sneutrino ``Majorana'' (lepton number violating)
mass $m^2_{\tilde{\nu}_3 \tilde{\nu}_3}$, and the
misalignment from off diagonal slepton masses
$\Delta_{k3}'$,
treated here in the mass insertion approximation. 
This diagram is in the tree level neutrino mass
eigenstate basis.}
\label{f5}
\end{figure}

We now proceed to discuss the loop contributions to the
neutrino masses which are zero at tree-level, again working
in the basis $\nu^i=(\nu'_e, \nu'_0, \nu_3)$ 
in which the tree-level neutrino masses are diagonal.
As discussed such corrections rely on flavour-violating effects,
and  non-zero masses develop at one loop
via the neutralino exchange diagram with flavour-violating mass
insertions along the sneutrino propagator. In the basis in which we
are working the only
non-zero Majorana sneutrino mass 
%in the absence of such insertions 
is $m_{{\tilde \nu}_3{\tilde \nu}_3}$, and so the diagram must always
have this single Majorana mass insertion at its centre. 
The flavour-violating mass insertions  therefore must
connect the neutrino of interest to ${\tilde \nu}_3$, and so the
mass insertions which are relevant are of the form $\Delta'_{i3}$.
See Fig.\ref{f5}. 

The one-loop corrected neutrino mass matrix,
to lowest non-zero order in the $\Delta'_{i3}$s, is given by:
\beq
m^{(1)}_{{\nu}_i{\nu}_j} =
\left( \begin{array}{lll}
{\Delta '}^2_{13} & \Delta'_{13} \Delta'_{23} & \Delta'_{13}     \\
\Delta'_{32} \Delta'_{31} & {\Delta '}^2_{23} & \Delta'_{23}\\
\Delta'_{31} &  \Delta'_{32} & 1
\end{array}
\right)\epsilon m_{{\nu}_3} +
\left( \begin{array}{lll}
0 & 0 & 0 \\
0 & 0 & 0 \\
0 & 0 & m_{{\nu}_3} 
\end{array}\right)
\label{Maj1}
\eeq
The upper $2\times 2$ block of the matrix clearly requires 
two mass insertions and so may be thought to be negligible
compared to the other elements of the matrix.
However on the contrary
this block actually gives the dominant contribution
to the neutrinos which are massless at tree-level.
The reason is that the contributions to these masses
from the third row and column of the matrix are
suppressed by a sort of see-saw mechanism.
Thus if one only includes $\Delta$ to
linear order, then the neutrinos which are massless at tree-level
will get see-saw
masses $\sim \Delta^2 \epsilon^2 m_{{\nu}_3} $,
which are to small to be interesting
(suppressed by two powers of $\epsilon$). 
Therefore to excellent approximation we may just consider the
upper $2\times 2$ block of the matrix involving two mass insertions.
In the basis $\nu^i=(\nu'_e, \nu'_0)$  this is of the form
\beq
m^{(1)}_{{\nu}_i{\nu}_j} =
\left( \begin{array}{ll}
{\Delta '}^2_{13} & \Delta'_{13} \Delta'_{23}      \\
\Delta'_{32} \Delta'_{31} & {\Delta '}^2_{23} 
\end{array}
\right)\epsilon m_{{\nu}_3} 
\label{Maj2}
\eeq
which has one zero eigenvalue $m_{{\nu}_1}=0$ and one non-zero eigenvalue 
\beq
m_{{\nu}_2}= ({\Delta'}^2_{13} + {\Delta'}^2_{23})\epsilon m_{{\nu}_3} 
\eeq
The corresponding eigenvectors are related to 
the tree-level massless eigenstates $\nu'_e, \nu'_0$ by
\beq
\left(
\begin{array}{l}
\nu_1 \\
\nu_2
\end{array}
\right)
=
\left(
\begin{array}{ll}
c_{12} & -s_{12}\\
s_{12} & c_{12}
\end{array}
\right)
\left(
\begin{array}{l}
\nu'_e \\
\nu'_0
\end{array}
\right)
\label{remix}
\eeq
where the mixing angle is given by
\beq
t_{12}=\Delta'_{13}/\Delta'_{23}
\eeq
Thus the SUSY radiative corrections induce a further two-state remixing of
$\nu'_e, \nu'_0$.
The final one-loop corrected neutrino mixing matrix 
is obtained from Eqns. \ref{defns},\ref{U0},\ref{remix},
\beq
\left(
\begin{array}{l}
\nu_e  \\
\nu_\mu  \\
\nu_\tau 
\end{array}
\right)_L = U'
\left(
\begin{array}{l}
\nu_1 \\
\nu_2 \\
\nu_3
\end{array}
\right)_L
\label{defns1}
\eeq
where
\beq
U' = \left( \begin{array}{ccc}
c_{12}-\frac{s_{12}}{t_{23}}\theta_1
&   s_{12}+\frac{c_{12}}{t_{23}} \theta_1   & \theta_1     \\
-s_{12}c_{23}-\frac{c_{12}}{s_{23}}\theta_1  & 
c_{12}c_{23}-\frac{s_{12}}{s_{23}}\theta_1 & s_{23}  \\
s_{12}s_{23}  &  -c_{12}s_{23}   & c_{23} 
\end{array}
\right)
\label{U'}
\eeq

In order to obtain ``just so'' vacuum oscillations we must be able to
achieve two things:

(i) We need the mass $m_{{\nu}_2}$ to be of order
$10^{-5}$ eV. For $\epsilon \simeq 10^{-3}$
and $m_{\nu_3} \simeq 5 \times 10^{-2}$ eV,
we need ${\Delta'}^2_{13} + {\Delta'}^2_{23}
\simeq .2$.

(ii) We need a large mixing angle $\theta_{12} \sim \pi/4$
which corresponds to $t_{12}=\Delta'_{13}/\Delta'_{23}\sim 1$.

Using Eq.\ref{Deltarelation} we can relate $\Delta'_{ij}$ to the
$\Delta_{\a \b}$ quantities in the charged lepton mass
eigenstate basis where the phenomenological bounds from lepton
flavour violating processes appear.
For example there is an extremely strong model independent limit from
$\mu \rightarrow e \gamma$ on $\Delta_{e \mu}<7.7 \times 10^{-3}$
\cite{S} (for 100 GeV sleptons). Since the limit on this quantity is
so much stronger than on the other entries, we may assume it is zero.
If we also set $\theta_1=0$ then we find
\bea
{\Delta'}_{13} & = & {\Delta'}_{31}=c_{23}\Delta_{e \tau} \nonumber \\
{\Delta'}_{23} & = & {\Delta'}_{32}=
s_{23}c_{23}\Delta_{\mu \mu}+(2c_{23}^2-1)\Delta_{\mu \tau}
-s_{23}c_{23}\Delta_{\tau \tau}
\eea
Assuming maximal mixing in the 23 sector, the condition for
maximal mixing in the 12 sector is then
\beq
\Delta_{\tau e} \simeq\frac{1}{\sqrt{2}}
(\Delta_{\mu \mu}-\Delta_{\tau \tau)}
\eeq
Assuming maximal mixing in the 12 and 23 sectors
($c_{12} = s_{12} = c_{23} = s_{23} = 1/\sqrt{2}$),
we find ${\Delta'}^2_{13} + {\Delta'}^2_{23} = \Delta_{\tau e}^2$.
To get   $m_{{\nu}_2}\sim m_{\nu_3} \epsilon  \Delta^2 \sim 10^{-5}$ eV, 
we need $\Delta_{e \tau} \sim .4$.
The experimental bound on $\Delta_{e \tau}$ is 29, so there
is phenomenologically nothing wrong with
such a large $\Delta$. It is also
(just) small enough for the mass insertion approximation 
to be applicable, and since here we only interested in the 
order of magnitude of the effects our approximations are adequate.
In any case the $\Delta$s may be reduced slightly if
$\epsilon$ is somewhat larger than we have estimated.

The mechanism described here is in fact basis independent.
A compact notation which illustrates this is briefly
developed in the Appendix.

Finally we briefly discuss the contribution of the second
source of flavour violation, coming from the trilinear parameters,
to the masses of $\nu'_e, \nu'_0$.
The basic effect comes from the fact that in the charged lepton
mass eigenstate basis the trilinear parameters 
associated with the Yukawa couplings $\lambda_e, \lambda_{\mu}, \lambda_{\tau}$
may be unequal $A_e \neq A_{\mu} \neq A_{\tau}$. This would imply that
in the basis in which the Yukawa couplings are diagonal
apart from the trilinear mass $A_3$ associated
with the coupling
$v_2A_3\lambda_3\tilde{\nu}_3\tilde{N}$
there will be further (small) trilinear parameters $a_1, a_2$
associated with the couplings
$v_2a_1\tilde{\nu}'_e\tilde{N}, v_2a_2\tilde{\nu}'_0\tilde{N}$.
These couplings generate small off-diagonal sneutrino Majorana masses
via the mechanism in Fig.\ref{f3}. However the Yukawa couplings
$\lambda_1, \lambda_2$ are zero in this basis so sneutrino
masses are only generated in the third row and column
of the sneutrino Majorana matrix. If the same were
true for the corresponding neutrino masses it would lead to negligible
contributions to the masses of $\nu'_e, \nu'_0$ due to the see-saw
type suppression discussed above. However 
if we focus on the relevant upper $2\times 2$ block of the neutrino
Majorana mass matrix we can see that there is a one-loop diagram similar to
Fig.\ref{f5} but now involving both an off-diagonal sneutrino 
Majorana mass $m_{{\tilde \nu}_\i{\tilde \nu}_3}^2$ {\em and} a mass insertion
$\Delta'_{3j}$. This will lead to additional contributions 
to the $2\times 2$ neutrino matrix which for simplicity
we have not included in our estimates.

To summarise, we have shown that both the atmospheric and solar
neutrino data may be accounted for by a remarkably simple
model: the MSSM with the addition of a single right-handed neutrino $N$.
In the $\lambda_e =0$ limit when $\theta_1 =0$ 
the physics of 
atmospheric and solar neutrinos can be described very simply.
From the point of view of atmospheric oscillations the mass splitting
between the two lightest neutrinos is too small to be important
and the physics is described by the two state mixing of Eq.\ref{bbasis},
maintaining the $\nu_{\mu} \leftrightarrow \nu_{\tau}$ tree-level prediction.
From the point of view of solar oscillations since $\lambda_e =0$
the electron neutrino contains no component of $\nu_3$ and so
the physics is described by
two state mixing in Eq.\ref{remix} (dropping the primes) induced by
the radiative corrections.
The neutrino mixing matrix in Eq.\ref{U'} becomes:
\beq
U = \left( \begin{array}{ccc}
c_{12} &   s_{12}   & 0     \\
-s_{12}c_{23}  & c_{12}c_{23} & s_{23}  \\
s_{12}s_{23}  &  -c_{12}s_{23}   & c_{23} 
\end{array}
\right)
\label{U}
\eeq
When $\theta_{12} = \theta_{23} =\pi/4$ it corresponds to bi-maximal mixing
in the $(\nu_e - \nu_0)$
and $(\nu_{\mu} - \nu_{\tau})$ sectors \cite{bimaximal1}.
More generally
at tree-level the spectrum is controlled by the 4 parameters
$\lambda_{e}, \lambda_{\mu}, \lambda_{\tau}, M$ where we choose
$\lambda_{e} \ll \lambda_{\mu} \approx \lambda_{\tau}$
to obtain $\theta_{23} \approx \pi /4$, and the eigenvalue
$m_{\nu_3}=\lambda_{\a}^2v_2^2/M \sim 5 \times 10^{-2} \ eV$
to obtain $\Delta m_{23}^2=m_{\nu_3}^2
\sim 2.5 \times 10^{-3} \ eV^2$ as suggested by the recent atmospheric
data. The effect of radiative corrections is model dependent since
it depends on the SUSY masses, and crucially on the soft flavour-violating
parameters. To illustrate the mechanism we have made some simple 
estimates based on flavour violation due to the soft scalar masses only.
We have seen that it is quite reasonable to obtain
$\Delta m_{12}^2 \sim 10^{-10} \ eV^2$ 
and a large mixing angle $\theta_{12}$ from such effects.
Thus the origin of the tiny neutrino mass 
$m_{\nu_2} \sim 10^{-5} \ eV$ required for ``just so'' neutrino oscillations
may simply be due to the SUSY corrections
arising from the atmospheric neutrino mass in the single right-handed
neutrino model.

\begin{center}
{\bf Acknowledgements}
\end{center}
SFK would like to thank CERN and Fermilab for their hospitality during
the initial and final stages of this work.

\begin{center}
{\bf Appendix}
\end{center}

In equation (\ref{Maj1}), we wrote the neutrino mass
matrix as an expansion in
$\epsilon$ (the loop parameter) and the
$\Delta$s. To linear order in $\Delta$ and
$\epsilon$, the mass matrix 
is a seesaw, with $\nu_3$ playing
the role of the heavy neutrino.  The massless-at-tree-level
neutrinos $\nu_0$ and $\nu_e$ acquire
seesaw masses of order $\Delta^2 \epsilon^2 m_{\nu_3}$,
which we neglect because $\epsilon \sim 10^{-3}$. 
This means that we only
need to consider the 
$O(\Delta^2 \epsilon)$ submatrix 
involving $\nu_e$ and $\nu_0$. This is of
the form
\beq
[m_{\nu}^{loop}]_{ij} = 
\frac{ \epsilon}{M  } 
 (\Delta_{ik}  m_{Dk} m_{Dl}  \Delta_{lj} )
\eeq
in an arbitrary basis.
Note  that $(\nu_3)_i \propto m_{Di}$;
ie the direction of the massive-at-tree-level
neutrino is in the direction of the
Dirac mass, so I can write
\beq
 m_{Dk} m_{Dl} = |m_D|^2 [ \nu_3 \nu_3^T ]_{kl}
\eeq
The $2 \times 2$ matrix we are
interested in is therefore
\beq
\epsilon m_{\nu_3}
\left[
\begin{array}{cc}
 (\nu_e \cdot \Delta \cdot \nu_3)^2 &  (\nu_e \cdot \Delta \cdot \nu_3)
  (\nu_3 \cdot \Delta \cdot \nu_0 ) \\
 (\nu_e \cdot \Delta \cdot \nu_3)
  (\nu_3 \cdot \Delta \cdot \nu_0 ) &  (\nu_0 \cdot \Delta \cdot \nu_3)^2 \\
\end{array}
\right]
\label{CC}
\eeq
in $(\nu_e, \nu_0)$ basis. But we know
$\nu_3$, $\nu_e$ and $\nu_0$ in the charged
lepton mass eigenstate basis, from
equation (\ref{bbasis}), so we 
can evaluate the matrix elements
in the charged lepton mass eigenstate
basis where we know the experimental
bounds on the $\Delta$s. This gives
us the previously discussed results.

\end{document}